\documentclass[journal,12pt,onecolumn]{IEEEtran}
\usepackage{setspace}
\usepackage{amsmath,amsfonts}
\usepackage{algorithmic}
\usepackage{algorithm}
\usepackage{array}
\usepackage[caption=false,font=normalsize,labelfont=sf,textfont=sf]{subfig}
\usepackage{textcomp}
\usepackage{stfloats}
\usepackage{url}
\usepackage{verbatim}
\usepackage{graphicx}
\usepackage{cite}
\usepackage{setspace}
\usepackage{pgfplots}
\pgfplotsset{compat=1.18}
\begin{document}
\raggedbottom
\onecolumn
\title{Integrated Sensing, Communication, and Powering (ISCAP) for IoT: A Joint Beamforming Design}

\author{Maryam Asadi Ahmadabadi, S. Mohammad Razavizadeh, and Vahid Jamali

\thanks{The work of Vahid Jamali was supported in part by the LOEWE Initiative, Hesse, Germany, within the emergenCITY Center [LOEWE/1/12/519/03/05.001(0016)/72].}}
\maketitle
\begin{abstract}
This paper studies Integrated Sensing, Communication, and Powering (ISCAP) as a novel framework designed to enhance Internet of Things (IoT) applications within sixth-generation wireless networks. In these applications, in addition to IoT devices requiring an energy supply and receiving information or control data to perform their tasks, the base station serving them must sense the devices and their environment to localize them, thereby improving data transmission and enabling simultaneous power delivery. In our multi-node ISCAP IoT system, we optimize base station beamforming alongside the receiver's power-splitting factor to maximize energy harvesting while adhering to strict communication and sensing constraints. To effectively tackle this non-convex optimization problem, we decompose it into three manageable subproblems and employ several techniques such as semidefinite relaxation and Rayleigh quotient methods to find an efficient solution. Simulation results demonstrate the effectiveness of the proposed design, highlighting performance trade-offs among sensing accuracy, communication reliability, and power transfer efficiency.
\end{abstract}
\begin{IEEEkeywords}
Integrated sensing and communication (ISAC), Simultaneous wireless information and power transfer (SWIPT), Internet of Things (IoT), Beamforming design.
\end{IEEEkeywords}

\section{Introduction}
Sixth-generation (6G) wireless networks are poised to support a vast array of Internet of Things (IoT) applications, integrating billions of low-power devices for sensing and communication tasks \cite{8869705}. These devices necessitate stable energy sources, while base stations (BS) must deliver control and localization services \cite{9606831}. A significant challenge arises from the reliance on battery-powered IoT devices, which often require frequent replacements, thereby impacting network reliability and lifespan. Wireless power transfer (WPT) has emerged as a viable solution, particularly when integrated with simultaneous wireless information and power transfer (SWIPT) systems \cite{7327131}. This integration allows a single signal to deliver both energy and data, leading to significant improvements in several areas, including spectral efficiency, energy consumption, time delay, and interference management \cite{6957150}.

On the other hand, in addition to addressing the power supply challenges of IoT devices, BSs servicing these devices must also possess the capability to sense both the devices and their surrounding environments. This sensing capability is pivotal for various IoT applications, including logistics, autonomous vehicles, and UAV tracking during emergency scenarios. By accurately localizing devices and analyzing environmental conditions, BSs can enable precise data transmission and efficient power transfer. These functionalities are essential for optimizing routing processes and mitigating risks such as accidents, collisions, and traffic congestion.
In this context, while integrated sensing and communication (ISAC) has been widely studied, the joint design of sensing, communication, and power remains relatively unexplored. This emerging field, referred to as Integrated Sensing, Communication, and Power Transfer (ISCAP), aims to address these challenges.
ISCAP technology not only reduces hardware costs and delays by enabling simultaneous execution of these three tasks, but it also enhances spectral efficiency \cite{10663809}, which is expected to become increasingly important with the continued development of communication technologies and IoT.

Limited studies have explored the performance trade-offs among sensing, communication, and power transfer (S-C-P), particularly regarding joint beamforming design in these systems using various approaches. For instance, \cite{9977463} presents a beamforming scheme aimed at minimizing beam pattern mismatch error, considering transmit power budget constraints and QoS requirements for information receivers and  energy receivers.
Further contributions include \cite{10556683} that studies ISCAP beamforming designs in MIMO framework, whereas \cite{13} focuses on secure ISCAP using extremely large antenna arrays and employs near-field beam focusing. Moreover, \cite{5} explores beamforming designs of ISCAP systems using reconfigurable intelligent surfaces (RIS). The work in \cite{10681491} presents a joint hybrid beamforming and dynamic on-off control algorithm to minimize the total power consumption of the BS. Lastly, \cite{14} examines a more complex multi-user and multi-antenna setting, utilizing joint beamforming design for information and sensing/energy and proposes a practical beam scanning scheme for sensing.
While the designs presented in the aforementioned works cover various scenarios within the ISCAP domain, each node is assumed to be interested in only one function among sensing, communication, and power transfer. However, in applications such as retail, logistics, industrial IoT (IIoT), healthcare monitoring, and smart cities, it might be required that the BS delivers all three functions to a single node, rather than distributing these tasks across three separate nodes, which to the best of the authors' knowledge has not been investigated in the literature, yet. 

In this study, we address these gaps by considering a multi-functional BS that transmits an ISCAP signal designed to conduct communication signal transmission, power transfer, and sensing tasks simultaneously for  IoT nodes. From the SWIPT perspective, each IoT node is equipped with a power-splitter receiver, while from the ISAC perspective, a minimum SINR requirement is considered for both sensing and communication functionalities. Specifically, we aim to maximize sum harvested energy (HE) at each node by jointly optimizing transmit and receive beamforming at the BS alongside power splitting (PS) ratios at the receiver nodes (i.e., IoT nodes). For a nearly realistic environment of the mentioned IoT applications, this maximization is performed under a Rician channel that both the unified communication and power transfer channel, as well as the sensing channel, are modeled as Rician channels. Notably, to the best of our knowledge, considering the sensing channel as a Rician channel has not been explored in the ISCAP literature and represents another innovative aspect of this study. 

The proposed joint beamforming optimization leads to a non-convex problem, which complicates the derivation of the global optimum solution.
To address this issue effectively, we propose an efficient alternating optimization (AO) framework that decomposes the main problem into three interdependent subproblems. The first subproblem is reformulated using semidefinite relaxation (SDR), the second is linearized to enable efficient computation, and the third is solved as a Rayleigh quotient problem with a closed-form solution. This structured approach simplifies the overall problem while ensuring effective convergence. Simulation results validate the proposed framework, demonstrating significant trade-offs among sensing accuracy, communication reliability, and power transfer efficiency under practical constraints. This work offers a comprehensive solution for ISCAP and contributes to advancing sustainable IoT ecosystems.

\textit{Notation:}  
$\mathbb{C}^{x \times y}$ denotes the space of $x \times y$ complex-valued matrices.  
For a square matrix $\mathbf{X}$, $\mathrm{tr}(\mathbf{X})$ and $\mathbf{X}^{-1}$ denote its trace and inverse, respectively, while $\mathbf{X} \succeq 0$ means that $\mathbf{X}$ is positive semi-definite. For an arbitrary matrix $\mathbf{S}$, $\mathbf{S}^H$ and $\mathbf{S}^T$ denote the conjugate transpose and transpose of $\mathbf{S}$, respectively.  
$\mathbf{I}$ denotes the identity matrix with appropriate dimensions.  
$\text{E}[\cdot]$ represents the statistical expectation.  
$\|\mathbf{x}\|$ is the Euclidean norm of a complex vector $\mathbf{x}$, and $|\mathbf{x}|$ is the absolute value of a complex scalar $x$. The notation \( \mathcal{CN}(\mathbf{x}, \mathbf{\Sigma}) \) represents the distribution of a circularly symmetric complex Gaussian (CSCG) random vector with mean $\mathbf{x}$
and covariance matrix \( \mathbf{\Sigma} \), where \( \sim \) indicates "is distributed as."
\section{System Model}
Fig. 1 depicts a multi-node ISCAP IoT system comprising a multi-functional BS and $K$ IoT nodes. The BS is equipped with a uniform linear array (ULA) of $ N_t $ transmit antennas and $N_r $ receive antennas, operating in full-duplex (FD) mode. The transmit antennas broadcast multi-functional ISCAP signal to each node, facilitating communication, power transfer, and sensing, while the receive antennas simultaneously capture echo signals necessary to complete the sensing process.

Each IoT node is assumed to have a single antenna and a receiver equipped with a power splitter, allowing for simultaneous energy harvesting (EH) and information reception. In the sensing process, each IoT node is modeled as a point target, with sensing conducted in a mono-static configuration \cite{10556683}. Additionally, EH at each node is modeled linearly \cite{6805330}.
\subsection{Transmitted signal by BS}
The BS transmits a multi-functional ISCAP signal $\mathbf{x} \in \mathbb{C}^{N_t \times 1}$ that simultaneously meets the requirements for power transfer and information transmission at the receiver, and reflection sensing at the BS. This approach simplifies system design by eliminating the need for separate signal coordination, reducing interference, and conserving resources compared to independent signal management. Furthermore, concurrent execution of all three functions enhances system efficiency. Assuming linear beamforming, ISCAP signal can be expressed as
\begin{equation}
    \mathbf{x} = \sum_{k=1}^K \mathbf{w}_k s_k,
\end{equation}
where $s_k$ denotes the transmitted data symbol for node $k$, and $\mathbf{w}_k \in \mathbb{C}^{N_t \times 1}$ is the corresponding transmit beamforming vector. It is assumed that the average energy of this symbol is unitary, i.e., $\text{E}[|s_k|^2] = 1$.
The transmitted beamforming vectors, constrained by the maximum power limit $P_{\text max}$ at the BS, are given by
\begin{equation}
\begin{array}{l}
\text{E}[\left\| \mathbf{x} \right\|^2] = \sum_{k=1}^K \, \left\| \mathbf{w}_k \right\|^2 \leq P_{\text{max}}.
\end{array}
\end{equation}
\subsection{Received signal by nodes}
The received signal at each node is given by
\begin{equation}
    y_k = \mathbf{h}_k^H \, \sum_{j=1}^K \mathbf{w}_j s_j + n_k,
\end{equation}
where $\mathbf{h}_k$ denotes the channel between the BS and $k$'th node, which is assumed to be perfectly known at the transmitter. In this context, $n_k \sim \mathcal{CN}(0, \sigma_{k_c}^2)$ represents the additive white Gaussian noise (AWGN) at the antenna receiver. It also includes clutter interference and the residual self-interference resulting from full-duplex operation, which are considered as part of the background noise \cite{10382465}.\\ 
\begin{figure}[tp]
    \centering
    \includegraphics[scale=0.8]{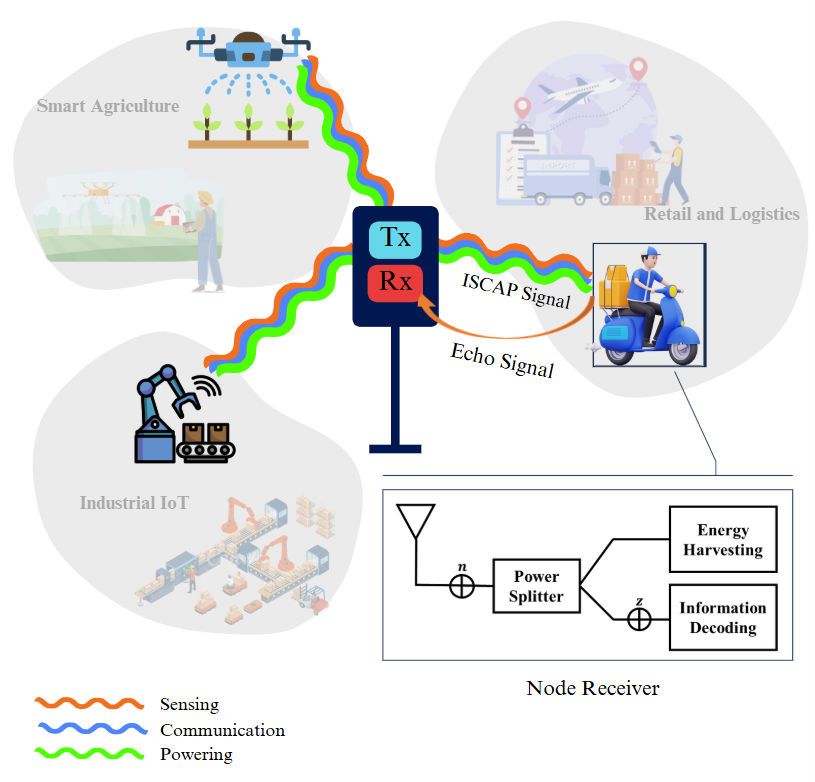}
    \caption{An IoT system equipped with ISCAP technology.}
    \label{SYSTEM1}
\end{figure}
To implement PS SWIPT technique, a power splitter is employed to divide the received RF power at the node into two parts: EH and information decoding (ID), using a splitting factor $\rho_k \in [0,1]$. With $\rho_k$ representing the power allocated to the EH part, the remaining fraction, $1 - \rho_k$, is allocated to the ID part.\\ 
The decoded signal at $k$'th node is expressed as \cite{6805330}
\begin{equation}
    y_k^{ID} = \sqrt{1-\rho_k} \, y_k + z_k,
\end{equation}
where $z_k \sim \mathcal{CN}(0, \delta_k^2)$ representing the extra noise added by the ID at the IoT node $k$. Consequently, the communication SINR at the ID part for node $k$ is given by
\begin{equation} \label{sinrc}
   \text{SINR}_k^{c}= \frac{(1-\rho_k)\, {| \mathbf{h}_k^H \, \mathbf{w}_k |}^2}{(1-\rho_k)\, \left( \sum_{j \ne k} {| \mathbf{h}_k^H \, \mathbf{w}_j |}^2 +  \sigma_{k_c}^2 \right) + \delta_k^2}.
\end{equation}
On the other hand, the signal allocated to EH is expressed as
\begin{equation}
    y_k^{EH} = \sqrt{\rho_k} \, y_k.
\end{equation}
Then, the harvested power based on the signal split to the EH at $k$'th node is given by
\begin{equation} \label{energy}
    \text{E}_k = \zeta_k \, \rho_k \, \left( \sum_{j=1}^K |\mathbf{h}_k^H \, \mathbf{w}_j|^2 + \sigma_{k_c}^2\right),
\end{equation}
where $\zeta_k \in (0,1]$ is the energy conversion efficiency of $k$'th node, which, for simplicity, is assumed to be 1.
\subsection{Received echo signals by BS}
After the ISCAP signal is transmitted by the BS, it simultaneously receives the echo signals from the nodes, denoted by $\mathbf{y}^{\text{BS}} \in \mathbb{C}^{N_r \times 1}$, which can be expressed as
\begin{equation}
    \mathbf{y}^{\text{BS}}=\sum _{k=1}^K \mathbf{H}_k \, \mathbf{x} + \mathbf{q} \, ,
\end{equation}
where $\mathbf{q} \sim \mathcal{CN}(0, \sigma_{s}^2) \in \mathbb{C}^{N_r \times 1}$ represents the receiver noise at the BS and $\mathbf{H}_k \in \mathbb{C}^{N_r \times N_t}$ denotes sensing channel for the $k$'th node. For $k$'th node after receive beamforming, the
received echo at its associated beam is given by
\begin{equation}
    y_k^{\text{BS}} = \mathbf{u}_k^H \mathbf{H}_k \mathbf{x} + q_k,
\end{equation}
where $\mathbf{u}_k \in \mathbb{C}^{N_r \times 1}$ is the receive beamforming vector associated with node $k$. 
To define an overall sensing objective related to the performance of various sensing tasks (such as detection, range/Doppler/angle estimation, and tracking), we consider sensing SINR as the performance metric, which is defined as \cite{10520715, 10742291}
\begin{equation}
    \begin{aligned} \label{sinrs}
        & \text{SINR}_k^{s}= \frac{\text{E} \{ | \mathbf{u}_k^H\, \mathbf{H}_k \mathbf{x}|^2 \}}{\sum_{j\ne k }\text{E} \{ | \mathbf{u}_k^H \mathbf{H}_j  \mathbf{x}|^2  \} +  \text{E} \{ |\mathbf{u}_k^H \mathbf{q}_k| \}}\\
        & \, \, \, \, \, \, \, \, \, \, \, \, \, \, \, \, \, \, \, \, = \frac{ \text{E} \{ \mathbf{u}_k^H  \mathbf{H}_k \mathbf{x} \mathbf{x}^H  \mathbf{H}_k^H \mathbf{u}_k \} }{\sum_{j\ne k }\text{E} \{ \mathbf{u}_k^H  \mathbf{H}_j \mathbf{x} \mathbf{x}^H  \mathbf{H}_j^H \mathbf{u}_k \} +  \text{E} \{ \mathbf{u}_k^H \mathbf{q}_k \mathbf{q}_k^{H} \mathbf{u}_k \}}\\
        & \, \, \, \, \, \, \, \, \, \, \, \, \, \, \, \, \, \, \, \, = \frac{  \mathbf{u}_k^H  \mathbf{H}_k  \sum_{l=1}^K \mathbf{w}_l \mathbf{w}_l^H \mathbf{H}_k^H \mathbf{u}_k }{\sum_{j\ne k} \mathbf{u}_k^H  \mathbf{H}_j  \sum_{l=1}^K \mathbf{w}_l \mathbf{w}_l^H   \mathbf{H}_j^H \mathbf{u}_k  +  \mathbf{u}_k^H \sigma_{k_s}^2 \mathbf{u}_k }.
    \end{aligned}
\end{equation}
\subsection{Channel Model}
As previously mentioned, we investigate the proposed system model under a Rician channel, in which both the unified communication and power transfer channels between the BS and the IoT nodes, as well as the sensing channel (originating from the BS transmit antennas, reflecting off an IoT node, and returning to the BS receiver antennas), are modeled as Rician channels. In this regard, the communication and sensing channels are modeled as follows.

a) \textit{Communication and powering channel:} The channel for communication and powering is given by
\begin{equation}
    \mathbf{h} = \sqrt{\frac{\kappa}{\kappa + 1}} \, \mathbf{h}_{\text{LoS}} + \sqrt{\frac{1}{\kappa + 1}} \, \mathbf{h}_{\text{NLoS}},
\end{equation}
where $\kappa$ is the Rician factor, $\mathbf{h}_{\text{LoS}} \in \mathbb{C}^{N_t \times 1}$ is the LOS deterministic component, and $\mathbf{h}_{\text{NLoS}} \in \mathbb{C}^{N_t \times 1}$ denotes the Rayleigh fading component, which is defined as follows
\begin{equation}
\mathbf{h}_{\text{LoS}} = \sqrt{L_0 \, \left(\frac{d_k}{d_0}\right)^{\alpha}} \, \mathbf{a}_t(\theta_k, \phi_k),
\end{equation}
where $L_0 = -40\, \mathrm{dB}$ represents the channel gain at the reference distance $d_0 = 1\, \mathrm{m}$, $d_k$ denoting the distance between each IoT node and the BS, and $\alpha$ denotes the path-loss exponent in the LoS case. Moreover, the NLoS component is modeled as
\begin{equation}
\mathbf{h}_{\text{NLoS}} = \sqrt{L_0 \, \left(\frac{d_k}{d_0}\right)^{\tilde{\alpha}}} \, \tilde{\mathbf{h}}_{\text{NLoS}},
\end{equation}
where $\tilde{\alpha}$ is the path-loss exponent in the NLoS case, and $\tilde{\mathbf{h}}_{\text{NLoS}}$ is an i.i.d. complex Gaussian vector, with each element follows $[ \mathbf{h}_{\text{NLoS}} ]_{i} \sim \mathcal{CN} (0,1)$.

b) \textit{Sensing channel:} The sensing channel for the mono-static mode is modeled as
\begin{equation}
    \mathbf{H} = \sqrt{\frac{\kappa}{\kappa + 1}} \, \mathbf{H}_{\text{LoS}} + \sqrt{\frac{1}{\kappa + 1}} \, \mathbf{H}_{\text{NLoS}}.
\end{equation}
For the LoS component, $\mathbf{H}_{\text{LoS}} \in \mathbb{C}^{N_r \times N_t}$,
\begin{equation}
\mathbf{H}_{\text{LoS}} = \sqrt{L_0 \, \left(\frac{2 \, d_k}{d_0}\right)^{\alpha} \, \nu_k } \, \mathbf{a}_r(\hat{\theta_k}, \hat{\phi_k}) \, \mathbf{a}_t(\theta_k, \phi_k)^H,
\end{equation}
where $\nu_k$ represents the  radar
cross section (RCS) of the $k$'th IoT node.
For the NLoS component caused by scattering paths, $\mathbf{H}_{\text{NLoS}} \in \mathbb{C}^{N_r \times N_t}$,
\begin{equation}
\mathbf{H}_{\text{NLoS}} = \sqrt{L_0 \, \left(\frac{2 \, d_k}{d_0}\right)^{\tilde{\alpha}} \, \nu_k} \, \tilde{\mathbf{H}}_{\text{NLoS}},
\end{equation}
where $\tilde{\mathbf{H}}_{\text{NLoS}}$ is an i.i.d. complex Gaussian matrix, with each element follows $[ \mathbf{H}_{\text{NLoS}} ]_{i,j} \sim \mathcal{CN} (0,1)$.\\
The vectors $\mathbf{a}_t(\cdot)$ and $\mathbf{a}_r(\cdot)$ denote the steering vectors of the transmit and receive antennas, respectively. The angles $(\theta_k, \phi_k)$ and $(\hat{\theta}_k, \hat{\phi}_k)$ correspond to the angle of departure (AoD) and angle of arrival (AoA) for the $k$-th node. Since we assume mono-static sensing, the AoA and AoD are equal, i.e., $(\hat{\theta}_k, \hat{\phi}_k) = (\theta_k, \phi_k)$, where $\theta_k$ and $\phi_k$ are the azimuth and elevation angles of the $k$'th node.

Assuming half-wavelength spacing between adjacent antennas and taking the center of the ULA as the reference point, the steering vectors are expressed as
\begin{equation}
\begin{array}{l}
\mathbf{a}_t(\theta_k, \phi_k) = \left[ {e^{-j \pi (N-1)  \beta_k}, \dots , e^{j \pi N_t \beta_k} } \right]^T \, \in \mathbb{C}^{(N_t \times 1)}
\end{array}
\end{equation}

\begin{equation}
\begin{array}{l}
\mathbf{a}_r( \hat{\theta_k}, \hat{\phi_k}) = \left[e^{j \pi (N_t + 1) \beta_k}, \dots , e^{j \pi (N - 1) \beta_k} \right]^T \, \in \mathbb{C}^{(N_r \times 1)}
\end{array}
\end{equation}
where $N = N_t + N_r$ represents the total number of transmit and receive antennas, and $\beta_k$ is defined as $\beta_k = \sin(\theta_k) \cos(\phi_k)$.

To represent the 3D Cartesian coordinates of the $k$-th IoT node and the BS, let $p_k^{\text{node}} = [\mathsf{x}_k, \mathsf{y}_k, \mathsf{z}_k]$ and $p^{\text{BS}} = [\mathsf{x}_{\text{BS}}, \mathsf{y}_{\text{BS}}, \mathsf{z}_{\text{BS}}]$, respectively. Then, $\sin(\theta_k)$ and $\cos(\phi_k)$ are calculated as 
\begin{equation}
\begin{array}{l}
\sin(\theta_k) = \frac{\mathsf{y}_{\text{BS}} - \mathsf{y}_k}{\sqrt{(\mathsf{x}_{\text{BS}} - \mathsf{x}_k)^2 + (\mathsf{y}_{\text{BS}} - \mathsf{y}_k)^2}},
\end{array}
\end{equation}

\begin{equation}
\begin{array}{l}
\cos(\phi_k) = \frac{\mathsf{z}_{\text{BS}} - \mathsf{z}_k}{\sqrt{(\mathsf{x}_{\text{BS}} - \mathsf{x}_k)^2 + (\mathsf{y}_{\text{BS}} - \mathsf{y}_k)^2 + (\mathsf{z}_{\text{BS}} - \mathsf{z}_k)^2}}.
\end{array}
\end{equation}
\section{Joint Design in ISCAP Systems}
The aim of this study is to jointly optimize the transmit and receive beamforming at the BS and the PS ratio at each node to maximize the sum HE across all nodes. This optimization is constrained by the requirement that communication SINR, sensing SINR, and harvested power, defined in equations (\ref{sinrc}), (\ref{sinrs}) and (\ref{energy}), respectively, meet the specified thresholds $\gamma_k$, $\eta_k$ and $e_k$ for each node $k$. Additionally, the solution must satisfy the BS's transmit power constraint. Formally, the optimization problem is expressed as
\begin{equation}
\begin{aligned}
\text{P0}:\max_{\{\mathbf{w}_{k}, \mathbf{u}_k, \rho_k \}} \quad  & \sum_{k=1}^K \, \text{E}_k\\
\textrm{s.t.}  \quad & \text{C}_1: \, \text{SINR}_k^{c}(\mathbf{w}_{k}, \rho_k ) \ge \gamma_{k} & \forall k\in K\\
&\text{C}_2: \, \text{SINR}_k^{s}(\mathbf{w}_{k}, \mathbf{u}_k) \ge \eta_{k} & \forall k\in K\\
&\text{C}_3: \, \text{E}_k(\mathbf{w}_{k}, \rho_k ) \ge e_k & \forall k\in K\\
&\text{C}_4: \, \sum_{k=1}^K {\| \mathbf{w}_{k} \|}^2 \leq P_{\text{max}} \\
&\text{C}_5: \, 0 < \rho_k < 1. & \forall k\in K
\label{p0}
\end{aligned}
\end{equation}   
Here, $P_{\text{max}}$ denotes the maximum allowable transmit power at the BS. Note that in this paper, we consider a general scenario where all nodes have non-zero SINR and harvested power targets, i.e., $\gamma_{k} > 0$ and $\text{E}_k > 0$ for all $k$. Therefore, the receive PS ratios at each node must satisfy $0 < \rho_k < 1$ for all $k$, as indicated by the final constraint in (\ref{p0}).\\
Due to the product terms involving the optimization variables, problem $\text{P0}$ is inherently non-convex. To address this, we decompose $\text {P0}$ into three subproblems. Using an AO approach, each subproblem is solved iteratively by optimizing one variable or a set of variables while keeping the others fixed, and this process is repeated until convergence is achieved.
\subsection{ Subproblem 1: Transmit Beamforming Optimization}
With fixed receive beamformers, ${\mathbf{u}_k}$, and PS ratio, $\rho_k$, where $k \in \{1, \dots, K\}$ problem $\text{P0}$ resembles a transmit beamforming optimization problem. 
\begin{equation}
\begin{aligned}
 \text{P1}:\max_{\{\mathbf{w}_{k} \}} \quad  & \sum_{k=1}^K \, \left( \sum_{j=1}^K |\mathbf{h}_k^H \, \mathbf{w}_j|^2 + \sigma_{k_c}^2\right)  \\
\textrm{s.t.}  \quad &\text C_1 \, , \text C_2 \,, \text C_3 \, , \text C_4 \,
\label{p}
\end{aligned}
\end{equation}
This problem is non-convex, therefore we transform it to SDP form, we define the matrix $\mathbf{W}_k = \mathbf{w}_k \, \mathbf{w}_k^H$, where $\mathbf{W}_k$ is a positive semidefinite (PSD) matrix with a rank one constraint, i.e., $\text{Rank}(\mathbf{W}_k)=1$. This problem is still non-convex. By utilizing SDR techniques to relax the non-convex rank one constraints, the resultant problem is formulated as
\begin{equation}
\begin{aligned}
\text{P2}: \max_{\{\mathbf{W}_k \}} \quad  & \sum_{k=1}^K \, \text{tr}(\mathbf{F}_k \, \mathbf{R}) \\
\textrm{s.t.}  \quad & \tilde{C}_1:\,   (1+ \gamma_k)\,  \text{tr}(\mathbf{F}_k \, \mathbf{W}_k) - \gamma_k\, \text{tr}(\mathbf{F}_k \, \mathbf{R}) & \ge \gamma_k \, \sigma_{k_c}^2 + \frac{\gamma_k \, \delta_k^2}{1 - \rho_k} \\
&\tilde{C}_2:\, (1+\eta_k) \, \text{tr}(\mathbf{G}_k \, \mathbf{R}) - \eta_k \, \sum_{j=1}^K \text{tr}(\mathbf{G}_j \, \mathbf{R}) & \ge \eta_{k}\, \sigma_{k_s}^2 \, \text{tr}(\mathbf{u}_k \, \mathbf{u}_k^H)\\
&\tilde{C}_3:\, \rho_k \, (\text{tr}(\mathbf{F}_k \, \mathbf{R}) + \sigma_{k_c}^2 ) \ge e_k \\
&\tilde{C}_4:\, \text{tr}(\mathbf{R}) \leq P_{\text{max}}\\
&\tilde{C}_5:\, \mathbf{W}_k \succeq 0,
\label{p2}
\end{aligned}
\end{equation}
where $\mathbf{F}_k=\mathbf{h}_k\, \mathbf{h}_k^H $, $\mathbf{g}_j=\mathbf{H}_j^H \, \mathbf{u}_k $, $\mathbf{G}_k=\mathbf{g}_k \, \mathbf{g}_k^H$, and $\mathbf{R}=\sum_{k=1}^K \mathbf{W}_k$. This problem is convex with respect to $\mathbf{W}$, which can be solved using standard solvers such as CVX.\\
The optimal solution of $\text P2$, denoted by \( \mathbf{W}^*_k \), is not necessarily guaranteed to have a rank of 1. If \( \mathbf{W}^*_k \) is rank-1, the unique eigenvalue corresponds to the trace of \( \mathbf{W}^*_k \), and the associated eigenvector defines the optimal beamforming vector. However, if the solution is not rank-1, a rank-1 solution must be obtained from \( \mathbf{W}^*_k \). A widely-adopted technique to extract a rank-1 solution is Gaussian Randomization (GR) \cite{8811733}, where a random vector \( \mathbf{r} \) is generated from \( \mathcal{CN}(0, \mathbf{W}^*_k) \) and subsequently projected onto the feasible region of $\text P2$. This random sampling process is repeated several times, and the best approximate solution is selected.
\subsection{ Subproblem 2: Power Splitting Optimization}
With fixed transmit and receive beamformers, for each $k \in \{1, \dots, K\}$, problem $\text{P0}$ reduces to an optimization of the PS ratio. 
\begin{equation}
\begin{aligned}
\text{P3}: \max_{\{\rho_k \}} \quad  &  \sum_{k=1}^K \,\rho_k \\
\textrm{s.t.}  \quad & \text{C}_1 \,, \text{C}_3 \, , \text{C}_5 \,
\label{p3}
\end{aligned}
\end{equation}
This problem, formulated in terms of $\rho_k$, is linear and is solved using e.g. CVX.
\subsection{ Subproblem 3: Receiver Beamforming Optimization}
For a given transmitted beamforming, problem $\text{P0}$ resembles a receive beamforming optimization problem. To support the indirect aim of power reduction, $\mathbf{u}_k$ is optimized to enhance the sensing SINR of each node. This method leverages the sensing SINR’s structure, transforming the subproblem into a generalized Rayleigh quotient problem with a known closed-form solution. Thus, we have the following optimization problem:
\begin{equation}
\begin{aligned}
\text{P4}: \max_{\{\mathbf{u}_k \}} \quad  &  \frac{  \mathbf{u}_k^H  \, (\mathbf{H}_k \, \mathbf{S} \, \mathbf{H}_k^H ) \, \mathbf{u}_k }{ \mathbf{u}_k^H \, (\sum_{j\ne k} \mathbf{H}_j \, \mathbf{S} \, \mathbf{H}_j^H  +  \sigma_{k_s}^2 \mathbf{I}_{N_r} ) \, \mathbf{u}_k },
\end{aligned}
\label{p4}
\end{equation}
where $\mathbf{S}=\sum_{l=1}^K \mathbf{w}_l \mathbf{w}_l^H $. The objective function in \text{P4} can be restated as the following optimization problem
\begin{equation}
\begin{aligned}
\text{P5}: \max_{\{\mathbf{u}_k \}} \quad  &  \frac{  \mathbf{u}_k^H  \, \mathbf{A} \, \mathbf{u}_k }{ \mathbf{u}_k^H \, \mathbf{B} \, \mathbf{u}_k },
\end{aligned}
\label{p3-1}
\end{equation}
where $\mathbf{A}=\mathbf{H}_k \mathbf{S} \mathbf{H}_k^H$ and $\mathbf{B} = \sum_{i=1, i\ne k }   \mathbf{H}_i \mathbf{S}  \mathbf{H}_i^H  +  \sigma_{k_s}^2 \mathbf{I}_{N_r} $.
Problem $\text{P5}$ is a generalized Rayleigh ratio quotient problem \cite{10520715}. The optimal received beamformer solution
is the eigenvector associated with the largest eigenvalue of the matrix obtained by $\mathbf{B}^{-1}\,\mathbf{A}$. This result is well-known in the theory of generalized eigenvalue problems.\\
To solve problem $\text P0$, we alternately tackle subproblems $\text{P2}$, $\text{P3}$, and $\text{P5}$ in an iterative process. The detailed steps of the proposed AO algorithm are outlined in Algorithm 1.
\begin{algorithm}[t]
\caption{Proposed Alternating Optimization Algorithm}
\label{alg:AO_algorithm}
\begin{algorithmic}[1]
\STATE \textbf{Initialize:} The initial receive beamforming $\mathbf{u}_k^{(1)}$, PS ratio $\rho_k^{(1)}$, iteration number $r = 1$, and tolerance value $\tau$.
\REPEAT
    \STATE Solve problem $(\text{P2})$ for a given $\mathbf{u}_k^{(1)}$ and $\rho_k^{(1)}$, then apply EVD or reconstruct the optimization variables to obtain the feasible solution denoted as $\mathbf{w}_k^{(r)}$.
    \STATE Solve the problem $(\text{P3})$ for a obtained $\mathbf{w}_k^{(r)}$.
    \STATE Solve the problem $(\text{P5})$ for a obtained $\mathbf{w}_k^{(r)}$.
    \STATE Update $r = r + 1$
\UNTIL {$|\, ( \sum_{k=1}^K \, E_k^{(r+1)}) - (\sum_{k=1}^K \, E_k^{(r)} \,| \leq \tau$}
\end{algorithmic}
\end{algorithm}
\subsection{Proposed Algorithm Analysis}
In the following, we study the initialization, convergence, feasibility and computational complexity of Algorithm 1.

\textit{1) Initialization}: 
This algorithm commences by initializing the combiners, $\mathbf{u}_k$, as matched filters for their respective steering vectors, assuming knowledge of the LoS angles. In each iteration, it iteratively refines the values of $\{ \mathbf{w}_k, \mathbf{\rho}_k , \mathbf{u}_k\}$ until the normalized improvement in the objective function falls below a threshold $\tau$.

\textit{2) Convergence}: The optimization problem is divided into three subproblems: transmit beamforming ($\mathbf{w}_k$), power splitting ratio ($\rho_k$), and receive beamforming ($\mathbf{u}_k$). Transmit beamforming is solved using convex relaxation (SDR) in CVX, power splitting is linear in $\rho_k$ and directly solvable via CVX, and receive beamforming has a closed-form solution as a generalized Rayleigh quotient problem. Due to the convexity of each subproblem, the objective value either increases or remains constant at each iteration.

The AO algorithm iteratively solves these subproblems, keeping the other variables fixed, to maximize the sum HE while satisfying constraints on communication SINR, sensing SINR, and power. The convergence of the AO algorithm is ensured by three key properties. First, the objective function $f$ (sum HE) does not decrease across iterations, a property known as monotonicity, $f(\mathbf{w}^{(r+1)}, \rho^{(r+1)}, \mathbf{u}^{(r+1)}) \geq f(\mathbf{w}^{(r)}, \rho^{(r)}, \mathbf{u}^{(r)})$. This is guaranteed because each subproblem optimizes a subset of the variables either exactly or approximately within numerical tolerances while keeping the others fixed. Second, the objective function $f(\mathbf{w}, \rho, \mathbf{u})$ is bounded above by a maximum feasible value $f_{\text{max}}$, determined by constraints such as the power limit ($P_{\text{max}}$). Finally, the sequence of objective values ${f^{(r)}}$ is monotonically non-decreasing and bounded, ensuring convergence to a finite value where $f^*$ represents the final converged value. However, our simulation results also support the validity of this point (Fig. 4).

\textit{3) Feasibility}: The optimization problem presented in (\ref{p0}) is not necessary always feasible; its feasibility is conditional, depending on the simultaneous satisfaction of all constraints, \( \text{C}_1\)–\( \text{C}_5\). 
To address the potential infeasibility of the problem, one strategy involves applying a constraint relaxation approach in cases where the problem becomes infeasible. Specifically, a decreasing sequence $\chi \in [0,1]$ is introduced, which scales the thresholds $\gamma_k$, $\eta_k$, and $e_k$ by $\chi$. This approach relaxes the constraints during the early iterations and gradually enforces them as the algorithm converges. By combining these strategies, the proposed algorithm effectively handles infeasibility challenges while enhancing convergence.

\textit{4) Complexity}: 
For the transmit beamforming optimization, which is solved via an SDP using the interior-point method, the complexity can be derived from \cite[Th. 3.12]{Pólik2010}. The computational complexity of an SDP problem with \( m \) constraints involving an \( n \times n \) PSD matrix is expressed as $\mathcal{O}\left(\sqrt{m} \log \frac{1}{\tau} \big(mn^3 + m^2n^2 + m^3\big)\right),$ where \( \tau > 0 \) represents the solution accuracy. By substituting \( n = N_t \) and \( m = 3K \), the approximate computational complexity for solving \( \text P2 \) becomes $\mathcal{O}\left(\sqrt{N_t} \log \frac{1}{\tau} \big(3K N_t^3\big)\right)$.
The second sub-problem, involving PS optimization, is formulated as a linear programming (LP) problem, with a computational complexity of \( \mathcal{O}(K^3) \) \cite{10.1145/800057.808695}.
For the final sub-problem, the optimal receive beamforming is determined using the Rayleigh quotient. Calculating the inverse of the matrix \( B \) for this purpose requires a complexity of \( \mathcal{O}(N_r^3) \) \cite{10520715}.
As a result, the overall complexity of the proposed sum HE maximization algorithm is $\mathcal{O}\left( N_{\text{iter}} \, \big( \sqrt{N_t} \log \frac{1}{\tau} \big(3K N_t^3\big) + K^3 + N_r^3 \big)\right)$, where $N_{\text{iter}}$ denotes the number of iteration of the Algorithm 1.

Therefore, the complexity of the sum HE maximization algorithm is determined not only by the number of iterations, denoted as $N_{\text{iter}}$, which is set to meet the convergence condition $\big| \big( \sum_{k=1}^K \text{E}_k^{(r+1)} \big) - \big( \sum_{k=1}^K \text{E}_k^{(r)} \big) \big| \leq \tau,$ but also by the complexities associated with the optimization of transmit beamforming, PS, and receive beamforming sub-problems.
\section{Simulation Results}
In this section, a numerical evaluation of the proposed beamforming design for ISCAP IoT systems is presented, considering two channel scenarios: LoS and Rician channels. The parameters for all nodes are assumed to be identical, and are given as $\delta_k^2 = \delta^2$, $\sigma_{k_c}^2 = \sigma_{k_s}^2 = \sigma^2$, $\gamma_k = \gamma$, $\eta_k = \eta$, $\nu_k = \nu$, and $e_k = e$, $\forall k$. The specific values used in the simulations are $\delta^2 = -100 \, \text{dBm}$, $\sigma^2 = -90 \, \text{dBm}$, $\gamma = \eta = 10 \, \text{dB}$, $\nu = 1 \, \text{m}^2$, $e = 1 \,\text{nW}$, $\tau = 10^{-8}$, $\kappa = 5 $, $\alpha = 2$, $\tilde{\alpha} = 3.2$, and the number of users is 4.
The BS is equipped with $N_t = 8$ transmit antennas and $N_r = 12$ receive antennas, and is positioned at a height of 10 meters. Additionally, the maximum transmit power of the BS is set to $P_{\text{max}} = 30 \, \text{dBm}$.

The subsequent sections provide a detailed analysis of the algorithm's performance and the tradeoff results for sensing, communication, and power transfer.
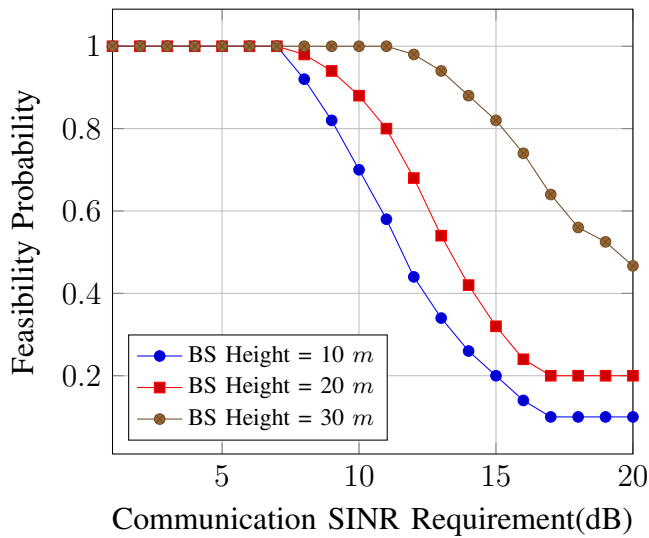
\begin{figure}
    \centering
    \begin{tikzpicture}
        \begin{axis}[
            width=8.5cm, height=7.5cm, 
            xlabel={Communication SINR Requirement(dB)},
            ylabel={Feasibility Probability},
            grid=major,
            xmin=1, xmax=20,
            legend style={
                legend pos= south west,
                font=\footnotesize, 
                cells={anchor=west}, 
            },
            legend entries={
                BS Height = 10 $m$, BS Height = 20 $m$, BS Height = 30 $m$ 
            }
        ]
        \addplot table [x index=0, y index=1] {data_node.txt}; 
        \addplot table [x index=0, y index=2] {data_node.txt};  
        \addplot table [x index=0, y index=3] {data_node.txt}; 
        \end{axis}
    \end{tikzpicture}
        \caption{Feasibility probability as a function of the number of IoT nodes for different BS heights in the LoS scenario.}
    \label{fig:feasibility}
\end{figure}
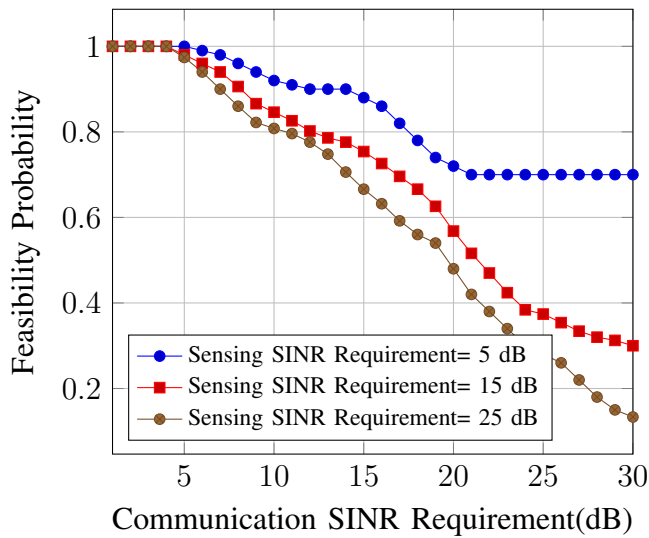
\begin{figure}
    \centering
    \begin{tikzpicture}
        \begin{axis}[
            width=8.5cm, height=7.5cm, 
            xlabel={Communication SINR Requirement(dB)},
            ylabel={Feasibility Probability},
            grid=major,
            xmin=1, xmax=30,
            legend style={
                legend pos= south west,
                font=\footnotesize, 
                cells={anchor=west}, 
            },
            legend entries={
                Sensing SINR Requirement= 5 dB, Sensing SINR Requirement= 15 dB , Sensing SINR Requirement= 25 dB
            }
        ]
        \addplot table [x index=0, y index=1] {data_C.txt}; 
        \addplot table [x index=0, y index=2] {data_C.txt};  
        \addplot table [x index=0, y index=3] {data_C.txt}; 

        \end{axis}
    \end{tikzpicture}
    \caption{Feasibility probability as a function of the communication SINR for different sensing SINR thresholds in a Rician channel scenario.}
    \label{fig:feasibility_rician}
\end{figure}
\begin{figure}
    \centering
    \begin{tikzpicture}
        \begin{axis}[
            width=8.5cm, height=7.5cm, 
            xlabel={Number of Iteration},
            ylabel={Sum Harvested Energy (W)},
            grid=major,
            xmin=1, xmax=10,
            legend style={
                font=\footnotesize, 
                cells={anchor=west}, 
                legend pos= south east,
            },
            legend entries={
                LoS Channel, $\kappa=10$, $\kappa=5$, $\kappa=1$ 
            }
        ]
        \addplot table [x index=0, y index=1] {data_conv.txt};  
        \addplot table [x index=0, y index=2] {data_conv.txt};  
        \addplot table [x index=0, y index=3] {data_conv.txt};  
        \addplot table [x index=0, y index=4] {data_conv.txt};  
        \end{axis}
    \end{tikzpicture}
\caption{Convergence behavior of sum HE across different $\kappa$.}
\label{fig:convergence}
\end{figure}
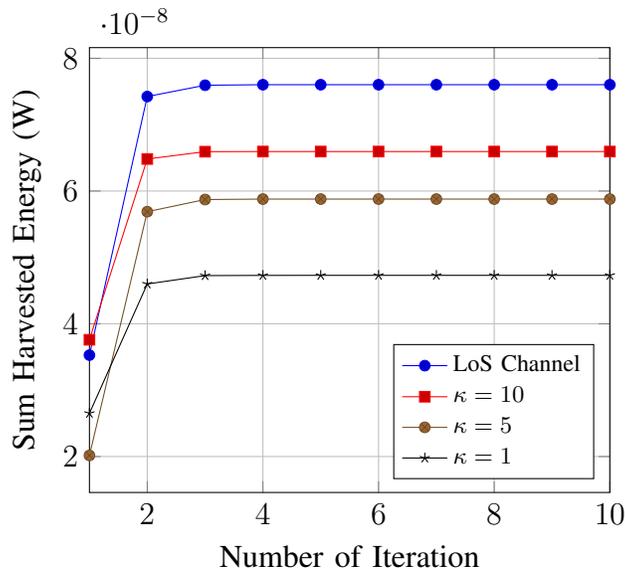
\subsection{Evaluation of the Algorithm's Performance}
Fig.~\ref{fig:feasibility} illustrates the feasibility probability as a function of the number of IoT nodes in the considered LoS scenario, IoT nodes are randomly distributed within a $50 \times 50 \, \text{m}^2$ area, with the BS located at the center of this region. The results are presented for three different BS heights. 

As observed, the feasibility probability decreases monotonically with the increasing number of IoT nodes. This behavior is attributed to the growing demand for limited system resources and the intensified interference as the density of IoT devices increases. Additionally, higher BS heights yield substantial improvements in feasibility probability. 
This observation reveals that increasing the BS height not only mitigates interference but also enhances signal propagation, enabling the system to sustain feasibility under higher user densities. Such findings provide actionable insights for optimizing network parameters, particularly in scenarios where the spatial density of IoT nodes is high and resource allocation becomes challenging.

Fig.~3 illustrates the feasibility probability as a function of the communication SINR for a Rician channel scenario. As mentioned, the users are randomly distributed across the area, while the channel incorporates both LoS and multipath components, accurately modeling realistic wireless environments. The results are depicted for three different sensing SINR thresholds. As observed, the feasibility probability decreases as the communication SINR increases across all sensing SINR thresholds. This trend reflects the increasing system constraints imposed by higher communication SINR requirements, which limit the available resources for meeting the sensing and EH demands. The results further underline the importance of joint resource allocation strategies in ISCAP systems. For instance, achieving higher feasibility probabilities under stringent sensing constraints necessitates a careful balance between communication, sensing, and power transfer. 

Fig.~\ref{fig:convergence} presents the convergence behavior of sum HE across different scenarios, including LoS and Rician channels with varying $\kappa$-factor. The LoS channel serves as an ideal reference, achieving the highest energy with rapid convergence, emphasizing the importance of the direct path for efficient energy transfer. As \( \kappa \) decreases, the impact of scattering increases, reducing harvested energy. For \( \kappa = 1 \), where scattering dominates, HE is minimized.

The optimization algorithm exhibits fast and stable convergence across all conditions, confirming its efficiency in complex environments. This analysis highlights the superiority of the LoS channel and the crucial role of the \( \kappa \)-factor in Rician channels.
\subsection{S-C-P Tradeoff and System Parameter Impact}
\begin{figure}
    \centering
    \begin{tikzpicture}
        \begin{axis}[
            width=8.5cm, height=7.5cm, 
            xlabel={Sensing SINR Requirement (dB)},
            ylabel={Sum Harvested Energy (W)},
            grid=major,
            xmin=1, xmax=10,
            legend style={
                font=\footnotesize, 
                cells={anchor=west}, 
                at={(0.98,0.98)}, anchor=north east 
            },
            legend entries={
                Comm. SINR Requirement = 5 dB, Comm. SINR Requirement = 10 dB, Comm. SINR Requirement = 15 dB,
            }
        ]
        \addplot table [x index=0, y index=1] {data_SEHSEn.txt};  
        \addplot table [x index=0, y index=2] {data_SEHSEn.txt};  
        \addplot table [x index=0, y index=3] {data_SEHSEn.txt};  
        \end{axis}
    \end{tikzpicture}
    \caption{The trade-off between S-C-P in the ISCAP system.}
    \label{fig:tradeoff}
\end{figure}
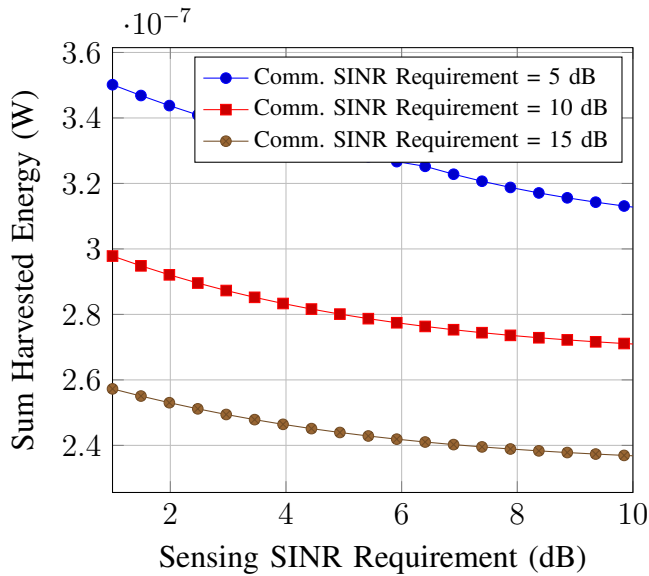
In Fig.~\ref{fig:tradeoff}, we study the trade-off between sensing SINR and sum HE in the ISCAP system. We observe from this figure that as the minimum sensing SINR increases, the sum HE decreases, highlighting the inherent challenge of resource allocation in such systems. This behavior occurs because the BS prioritizes sensing accuracy by allocating more beamforming gain to enhance the reception of reflected signals, which is crucial for accurate sensing. However, this reallocation of resources reduces the power available for EH, resulting in lower sum HE. Higher communication SINR imposes additional constraints on EH, highlighting the trade-offs in ISCAP system design for jointly optimizing sensing, communication, and EH.
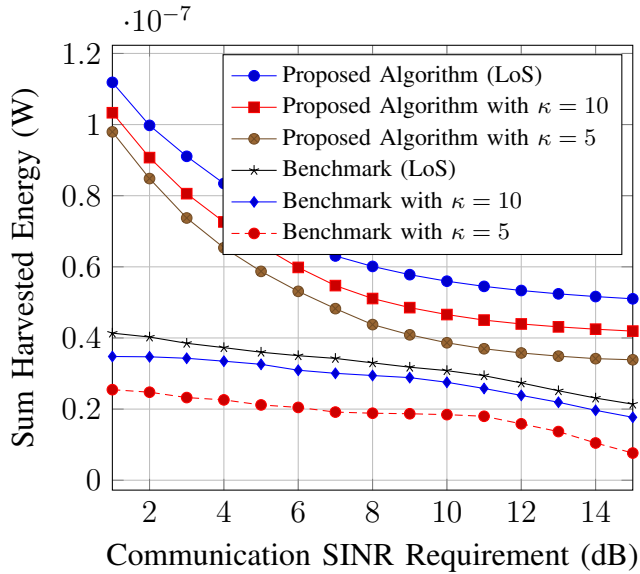
\begin{figure}
    \centering
    \begin{tikzpicture}
        \begin{axis}[
            width=8.5cm, height=7.5cm, 
            xlabel={Communication SINR Requirement (dB)},
            ylabel={Sum Harvested Energy (W)},
            grid=major,
            xmin=1, xmax=15,
            legend style={
                font=\footnotesize, 
                cells={anchor=west}, 
                at={(0.98,0.98)}, anchor=north east 
            },
            legend entries={
                Proposed Algorithm (LoS), Proposed Algorithm with $\kappa=10$, Proposed Algorithm with $\kappa=5$, 
                Benchmark (LoS), Benchmark with $\kappa=10$, Benchmark with $\kappa=5$
            }
        ]
        \addplot table [x index=0, y index=1] {data_EC.txt};  
        \addplot table [x index=0, y index=2] {data_EC.txt};  
        \addplot table [x index=0, y index=3] {data_EC.txt};  
        \addplot table [x index=0, y index=4] {data_EC.txt};  
        \addplot table [x index=0, y index=5] {data_EC.txt};  
        \addplot table [x index=0, y index=6] {data_EC.txt};  
        \end{axis}
    \end{tikzpicture}
        \caption{Comparison of the proposed algorithm and MRT in terms of EH under different $\kappa$.}
    \label{fig:6}
\end{figure}
Fig.~\ref{fig:6} presents a comparative analysis between the proposed algorithm and Maximum Ratio Transmission (MRT) under various channel conditions. 

The beamforming vector for MRT is defined as \(\mathbf{w}_k = \frac{\mathbf{h}_k}{\|\mathbf{h}_k\|}\) \cite{6783665}. Substituting this into the optimization framework, the optimal \(\rho_k\) is obtained by maximizing the objective function while satisfying constraints. Specifically, \(\rho_k\) is bounded by the minimum SINR and EH requirements $\rho_k \leq 1 - \frac{\gamma_k \delta_k^2}{ b - \gamma_k c}, \quad \rho_k \geq \frac{e_k}{ c + \sigma_{k_c}^2 }$.
Additionally, feasibility requires \(0 < \rho_k < 1\), leading to the optimal value $\rho_k^* = \max \left( \min \left( 1 - \frac{\gamma_k \delta_k^2}{b - \gamma_k c} , 1 \right), \frac{e_k}{ c } \right),$
where \( b=\frac{| \mathbf{h}_k^H \, \mathbf{h}_k |^2}{|\mathbf{h}_k|^2} \) and \( c=\sum_{j \ne k} \frac{| \mathbf{h}_k^H \, \mathbf{h}_j |^2}{|\mathbf{h}_j|^2} +  \sigma_{k_c}^2 \).  

In a LoS scenario, MRT steers the signal directly to the receiver but does not optimize power distribution across antennas. Advanced beamforming methods enhance channel utilization by efficiently allocating power, making MRT suboptimal.  

In Rician fading, the \(\kappa\)-factor influences performance. For \(\kappa = 5\), increased scattering degrades EH, with a noticeable decline as SINR rises, highlighting the SWIPT trade-off. Nonetheless, the proposed algorithm effectively balances this trade-off and consistently outperforms MRT across all SINR levels.  
\section{Conclusion}
In this paper, we investigate a multi-node ISCAP IoT system operating over Rician channels, where each node simultaneously handles information transmission, wireless power transfer, and sensing. The beamforming design is optimized to maximize the sum HE while ensuring that both sensing and communication SINR constraints are met. Simulation results validate the proposed scheme, demonstrating a balanced trade-off between sensing accuracy, communication reliability, and power transfer efficiency.
This work contributes to the growing research on ISCAP for IoT networks. While the current design employs mono-static sensing, future work will explore cooperative beamforming strategies with multiple BSs, as well as multi-static sensing approaches to enhance system performance.

\pagestyle{empty}
{

\bibliographystyle{ieeetr}
\onehalfspacing
\bibliography{Paper3_References}

\begin{thebibliography}{10}

\bibitem{8869705}
W.~Saad, M.~Bennis, and M.~Chen, ``{A Vision of 6G Wireless Systems: Applications, Trends, Technologies, and Open Research Problems},'' {\em IEEE Network}, vol.~34, no.~3, pp.~134--142, 2020.

\bibitem{9606831}
Y.~Cui, F.~Liu, X.~Jing, and J.~Mu, ``{Integrating Sensing and Communications for Ubiquitous IoT: Applications, Trends, and Challenges},'' {\em IEEE Network}, vol.~35, no.~5, pp.~158--167, 2021.

\bibitem{7327131}
X.~Lu, P.~Wang, D.~Niyato, D.~I. Kim, and Z.~Han, ``{Wireless Charging Technologies: Fundamentals, Standards, and Network Applications},'' {\em IEEE Communications Surveys \& Tutorials}, vol.~18, no.~2, pp.~1413--1452, 2016.

\bibitem{6957150}
I.~Krikidis, S.~Timotheou, S.~Nikolaou, G.~Zheng, D.~W.~K. Ng, and R.~Schober, ``{Simultaneous Wireless Information and Power Transfer in Modern Communication Systems},'' {\em IEEE Communications Magazine}, vol.~52, no.~11, pp.~104--110, 2014.

\bibitem{10663809}
X.~Li, Z.~Han, G.~Zhu, Y.~Shi, J.~Xu, Y.~Gong, Q.~Zhang, K.~Huang, and K.~B. Letaief, ``{Integrating Sensing, Communication, and Power Transfer: From Theory to Practice},'' {\em IEEE Communications Magazine}, vol.~62, no.~9, pp.~122--127, 2024.

\bibitem{9977463}
X.~Zeng, L.~Xing, Y.~Wu, and Y.~Shi, ``{Beamforming Design for Integrated Sensing and SWIPT System},'' in {\em 2022 IEEE 33rd Annual International Symposium on Personal, Indoor and Mobile Radio Communications (PIMRC)}, pp.~403--408, 2022.

\bibitem{10556683}
Z.~Zhou, X.~Li, G.~Zhu, J.~Xu, K.~Huang, and S.~Cui, ``{Integrating Sensing, Communication, and Power Transfer: Multiuser Beamforming Design},'' {\em IEEE Journal on Selected Areas in Communications}, vol.~42, no.~9, pp.~2228--2242, 2024.

\bibitem{13}
Z.~Ren, S.~Zhang, X.~Li, L.~Qiu, J.~Xu, and D.~W.~K. Ng, ``{Secure Communications in Near-Filed ISCAP Systems with Extremely Large-Scale Antenna Arrays},'' {\em arXiv preprint arXiv:2405.13634}, 2024.

\bibitem{5}
Y.~Yang, H.~Gao, X.~Yang, R.~Cao, and Y.~Fan, ``{Joint Beamforming for RIS-assisted Integrated Communication, Sensing and Power Transfer Systems},'' {\em IEEE Wireless Communications Letters}, 2023.

\bibitem{10681491}
Z.~Hao, Y.~Fang, X.~Yu, J.~Xu, L.~Qiu, L.~Xu, and S.~Cui, ``{Energy-Efficient Hybrid Beamforming with Dynamic On-Off Control for Integrated Sensing, Communications, and Powering},'' {\em IEEE Transactions on Communications}, pp.~1--1, 2024.

\bibitem{14}
Y.~Chen, C.~Hu, Z.~Ren, H.~Hu, J.~Xu, L.~Xu, L.~Liu, and S.~Cui, ``{Integrated Sensing, Communication, and Powering over Multi-antenna OFDM Systems},'' {\em arXiv preprint arXiv:2408.14156}, 2024.

\bibitem{6805330}
Q.~Shi, L.~Liu, W.~Xu, and R.~Zhang, ``{Joint Transmit Beamforming and Receive Power Splitting for MISO SWIPT Systems},'' {\em IEEE Transactions on Wireless Communications}, vol.~13, no.~6, pp.~3269--3280, 2014.

\bibitem{10382465}
Y.~Chen, H.~Hua, J.~Xu, and D.~W.~K. Ng, ``{ISAC Meets SWIPT: Multi-Functional Wireless Systems Integrating Sensing, Communication, and Powering},'' {\em IEEE Transactions on Wireless Communications}, vol.~23, no.~8, pp.~8264--8280, 2024.

\bibitem{10520715}
D.~Galappaththige, S.~Zargari, C.~Tellambura, and G.~Y. Li, ``{Near-Field ISAC: Beamforming for Multi-Target Detection},'' {\em IEEE Wireless Communications Letters}, vol.~13, no.~7, pp.~1938--1942, 2024.

\bibitem{10742291}
U.~Demirhan and A.~Alkhateeb, ``{Cell-Free ISAC MIMO Systems: Joint Sensing and Communication Beamforming},'' {\em IEEE Transactions on Communications}, pp.~1--1, 2024.

\bibitem{8811733}
Q.~Wu and R.~Zhang, ``{Intelligent Reflecting Surface Enhanced Wireless Network via Joint Active and Passive Beamforming},'' {\em IEEE Transactions on Wireless Communications}, vol.~18, no.~11, pp.~5394--5409, 2019.

\bibitem{Pólik2010}
I.~P{\'o}lik and T.~Terlaky, {\em Interior Point Methods for Nonlinear Optimization}, pp.~215--276.
\newblock Berlin, Heidelberg: Springer Berlin Heidelberg, 2010.

\bibitem{10.1145/800057.808695}
N.~Karmarkar, ``{A New Polynomial-Time Algorithm for Linear Programming},'' in {\em Proceedings of the Sixteenth Annual ACM Symposium on Theory of Computing}, STOC '84, (New York, NY, USA), p.~302–311, Association for Computing Machinery, 1984.

\bibitem{6783665}
S.~Timotheou, I.~Krikidis, G.~Zheng, and B.~Ottersten, ``{Beamforming for MISO Interference Channels with QoS and RF Energy Transfer},'' {\em IEEE Transactions on Wireless Communications}, vol.~13, no.~5, pp.~2646--2658, 2014.

\end{thebibliography}

} 

\vfill

\end{document}